\documentclass[12pt]{iopart}
\usepackage{graphicx}
\usepackage{epsfig}

\begin{document}

\title[Flux front propagation in a drilled HTS during a pulsed-field magnetization]{Pulsed-field magnetization of drilled bulk high-temperature superconductors: flux front propagation in the volume and on the surface}
\author{Gregory P~Lousberg$^{1,2,7}$, J-F~Fagnard$^{1}$, E~Haanappel$^{3,4}$, X~Chaud$^{5}$, M~Ausloos$^{6}$, B~Vanderheyden$^{1}$, and Ph~Vanderbemden$^{1}$}
\address{$^1$ SUPRATECS Research Group, Dept. of Electrical Engineering and Computer Science (B28), University of Li\`ege, Belgium}
\address{$^2$ Laboratoire National des Champs Magn\'etiques Intenses (LNCMI), Toulouse, France}
\address{$^3$ IPBS - UPS/CNRS UMR5089, Toulouse, France}
\address{$^4$ Universit\'e de Toulouse, UPS, IPBS, Toulouse, France }
\address{$^5$ CNRS, CRETA, Grenoble, France}
\address{$^6$ SUPRATECS (B5a), University of Li\`ege, Belgium}
\address{$^7$ FRS-FNRS fellowship}
\ead{gregory.lousberg@ulg.ac.be}

\begin{abstract}
\linespread{2}
   \selectfont
We present a method for characterizing the propagation of the magnetic flux in an artificially drilled bulk high-temperature superconductor (HTS) during a pulsed-field magnetization. As the magnetic pulse penetrates the cylindrical sample, the magnetic flux density is measured simultaneously in $16$ holes by means of microcoils that are placed across the median plane, {\it i.e.} at an equal distance from the top and bottom surfaces, and close to the surface of the sample. We discuss the time evolution of the magnetic flux density in the holes during a pulse and measure the time taken by the external magnetic flux to reach each hole. Our data show that the flux front moves faster in the median plane than on the surface when penetrating the sample edge; it then proceeds faster along the surface than in the bulk as it penetrates the sample further. Once the pulse is over, the trapped flux density inside the central hole is found to be about twice as large in the median plane than on the surface. This ratio is confirmed by modelling.

\end{abstract}

\submitto{\SUST}


\maketitle
\linespread{2}
   \selectfont 
\section{Introduction}

Bulk high-temperature superconductors (HTS) consisting of a single domain of (RE)BaCuO (RE denotes a rare earth) are able to trap very large magnetic fields~\cite{Trapped1,Trapped2,Trapped3,Murakami}. They are considered as promising substitutes for conventional ferromagnetic permanent magnets that are used in a series of industrial applications, such as in motors~\cite{Motor} and magnetic bearing systems~\cite{Flywheel, Maglev}.  

The HTS trapped field magnets can be activated by means of a pulsed magnetic field provided by the fast discharge of a capacitor through a copper coil. Because of the compactness of the exciting coil, the pulsed field magnetization technique is suitable for \textit{in situ} activation of the magnets, contrary to the field-cooled activation process that requires large and bulky superconducting magnets. 

When a HTS bulk magnet is activated by a pulsed field, the fast motion of vortices may locally produce a large amount of heat that strongly affects the magnetic flux propagation and the trapping properties of the HTS. It is thus of prime importance to characterize the penetration of the magnetic flux  in a bulk HTS magnet activated by a pulsed field. To this aim, many experiments involve either (i) the acquisition of the waveform $B(t)$  during the pulse with Hall probes glued on the surface of the sample~\cite{Chaud, Huo, Tokuyama}, or (ii) a Hall probe mapping realized after the application of the pulse~\cite{Tokuyama,Yokoyama, Tsuchimoto}. Although such mapping yields a complete distribution of the trapped flux above the sample surface with a good spatial resolution, it does not allow one to observe the flux penetration during the pulse. Conversely, the characterization technique (i) --- which involves Hall probes glued on the surface --- overcomes this limitation, but as the number of probes is limited (usually a maximum of six Hall probes are used~\cite{Huo, Tokuyama}) this method focuses on specific locations at the sample surface and fails to provide a detailed picture of the penetration of the magnetic flux on the surface. In a recent study~\cite{Shiraishi}, an attempt at refining the analysis of the flux penetration was carried by placing $25$ pick-up coils at $0.5~\mathrm{mm}$ above the surface of a pulse-field activated sample.

Overall, the conventional Hall probe methods enable one to determine the flux density distribution above the top surface of the sample but they do not provide direct information on the flux penetration and/or trapping in the bulk of the superconductor. An interesting step toward this direction was carried by Ikuta \textit{et al.} in Ref.~\cite{Ikuta}, who placed seven pick-up coils in the gap between two HTS bulk magnets arranged on top of each other, in order to estimate the flux that would be trapped in the median plane of a larger sample.

Recently, bulk HTS cylinders with artificial columnar holes have been synthesized; the presence of holes has been shown to enhance the chemical, thermal, and mechanical properties of the bulk HTS material~\cite{Murakami,H2,H3,H4}. Drilled structures also offer a means for characterizing locally the magnetic properties inside the volume of the magnet. The procedure is described in a previous work for the case of an AC magnetic field excitation~\cite{Fluxinside} and is based on the acquisition of the pick-up voltage across microcoils inserted \emph{inside} the holes, in order to probe the magnetic flux in the bulk of the sample.

In this paper, we use microcoils to locally measure the magnetic flux density inside the holes of a drilled HTS bulk magnets during a pulsed-field activation. In particular, we characterize the flux propagation in the median plane of the sample and compare it with its propagation near the surface. This non-destructive characterization technique also enables us to measure the trapped magnetic flux density in the median plane of the sample. Such pieces of information have not been available so far with traditional characterization techniques.

The paper is organized as follows. After a brief description of the measurement system in Section~\ref{s:Exp}, we present and discuss the experimental results in Section~\ref{s:results}. We analyse the time evolution of the magnetic flux density in the holes in Section~\ref{s:Time}. In Section~\ref{s:Fluxpropagation}, we characterize the flux penetration in the sample and determine the time taken by the applied field to penetrate a hole, either in the central region of the hole or near the surface. We discuss the distribution of the trapped flux density in Section~\ref{s:Trapped}. Section~\ref{s:Conclusion} presents our conclusions.

\section{Experimental setup}
\label{s:Exp}

\subsection{Bulk superconducting sample}
\label{ss:sample}

We used a top-seeded melt-grown single domain of YBa$_2$Cu$_3$O$_{7-\delta}$, synthesized at CRETA (Grenoble-France). The sample is a cylinder with a diameter of $16~\mathrm{mm}$ and a thickness of $10~\mathrm{mm}$. It contains 55 holes with a diameter of $0.8~\mathrm{mm}$. 

\begin{figure}[t]
\center
\includegraphics[width=10cm]{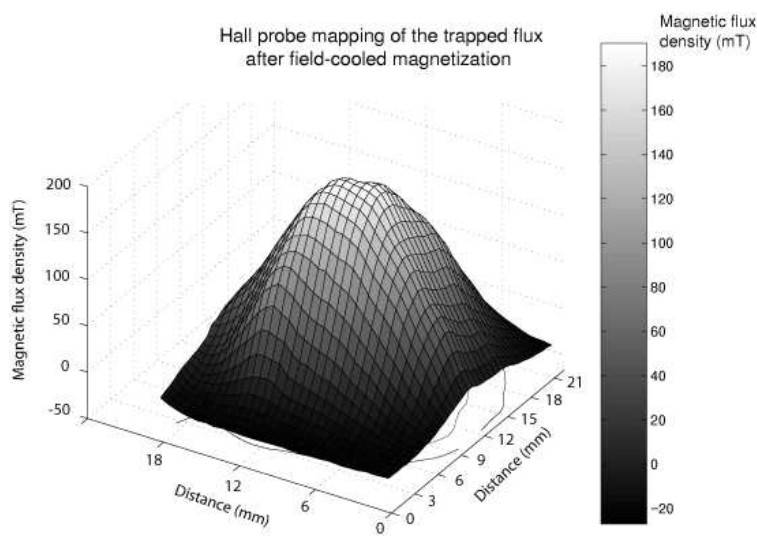}\caption{Hall probe mapping of the trapped magnetic flux density above the top surface of the drilled sample. The mapping is carried out $20~\mathrm{min}$ after a field-cooled process at $77~\mathrm{K}$ in a uniform magnetic flux density of $2~\mathrm{T}$.}\label{Mapping}
\end{figure}

The magnetic properties of the sample were first investigated after a field-cooled magnetization at $77~\mathrm{K}$ in a magnetic flux density of $2~\mathrm{T}$ provided by a superconducting coil. Once permanently magnetized, a two-dimensional mapping of the trapped magnetic flux density was performed about $20$ minutes after magnetization so as to reduce the effects of magnetic relaxation. The surface scan of the remnant magnetic flux density was made by displacing a Hall probe by steps of $0.5~\mathrm{mm}$ at $0.2~\mathrm{mm}$ above the top surface of the sample. The distribution of the trapped magnetic flux density, $B$, as measured above the surface, is shown in Figure~\ref{Mapping}. The axisymmetric trapped flux profile exhibits a single maximum, $B_{\mathrm{max}}=190~\mathrm{mT}$, and does not indicate the presence of macrocracks on the sample surface. Because of the distance between the probe and the sample surface, the flux distribution is smooth and the holes cannot be identified in the flux profile. The distribution observed in Figure~\ref{Mapping} is similar to  distributions observed previously in other drilled samples~\cite{H2,H3,H4}.

\subsection{Measurement technique}

The pulsed field magnetizations were carried out at the high magnetic field facilities of the ``Laboratoire National des Champs Magn\'etiques Intenses'' (LNCMI) on the Toulouse site (France). Pulsed fields with an amplitude of $3~\mathrm{T}$, a rising time of $60~\mathrm{ms}$, and a duration of $370~\mathrm{ms}$ were applied to the sample initially cooled down to $77~\mathrm{K}$ in the absence of magnetic field.

\begin{figure}[t]
\center
\includegraphics[width=8cm]{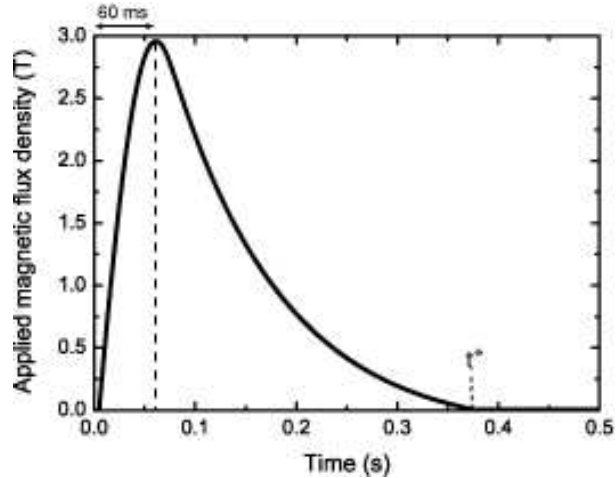}\caption{Applied magnetic pulse, as measured by a Hall probe in the cryostat containing no HTS sample.}\label{FigurePulse}
\end{figure}

The pulse of magnetic field was generated by a fast discharge current flowing through a copper coil placed in a cryostat. The current was produced by the discharge of a bench of capacitors ($470~\mathrm{mF}$) initially loaded at $1~\mathrm{kV}$.  The shape of the magnetic pulse was determined with a Hall sensor (Arepoc LHP-MU) placed in the experimental chamber inside the cryostat in the absence of the sample. The pulse is represented in Figure~\ref{FigurePulse}. It starts with a sharp increase of the magnetic flux density $B$, at a rate of $dB/dt\sim 100~\mathrm{T/s}$. The magnetic field then reaches a maximum of $60~\mathrm{ms}$, decays, and vanishes at $t^*=370~\mathrm{ms}$ when the electric circuit is open.

The sample was inserted in the cryostat with the help of a glass fiber rod terminated by a sample holder (also in glass fiber). The temperature of the sample ($77~\mathrm{K}$) was monitored by a LakeShore $340$ temperature controller.

\begin{figure}[t]
\center
\includegraphics[width=6cm]{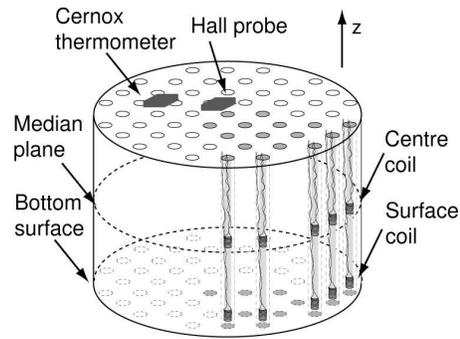}\caption{Schematic illustration of the drilled samples showing the $16$ holes (grey) where the magnetic flux density is measured. The coil holders, containing each two microcoils (a centre coil and a surface coil), are also represented in five peripheral holes.}\label{Sample}
\end{figure}

The magnetic flux threading the holes of the YBCO sample was probed by miniature coils. Each probed hole contained two coils, which  were wound on a Teflon holder tube with a height of $10~\mathrm{mm}$ and a diameter of $0.5~\mathrm{mm}$. The magnetic flux was probed in a quarter of the sample, containing a total of 16 holes  (these holes are represented in grey on Figure~\ref{Sample}). Note that earlier measurements with an AC magnetic field excitation (with a coil-probing method similar to that reported in our previous work~\cite{Fluxinside}) showed that this particular quarter of the sample is representative of the full sample. Each holder contained two microcoils consisting each of $10$ turns of $50~\mu\mathrm{m}$ of isolated copper wire, yielding a coil height of $1~\mathrm{mm}$. One coil was placed in the median plane of the sample at an equal distance from the top and bottom surfaces --- we will refer to it as the centre coil --- ; the other coil was put close to the sample surface --- we will refer to it as the surface coil ---. The axis of the coils was chosen to probe the vertical component of the induction ($B_z$), {\it i.e.} the component parallel to both the $c$-axis of the single domain and the direction of the columnar holes (see Figure~\ref{Sample}). The microcoils were connected to the measuring devices with finely twisted $50~\mu\mathrm{m}$ wires. A sketch of the microcoils is shown in Figure~\ref{Sample}. 

The induced electromotive force (emf) across each microcoil was recorded during the pulse. The signals were amplified by instrumentation amplifiers (INA 128 - gain of 10 000 and bandwidth of 10kHz) before being recorded for $500~\mathrm{ms}$ by a data acquisition card (National Instruments CompactDAQ - NI9215) at a high sampling rate ($100~\mathrm{kHz}$). The $16$ channels were acquired simultaneously so that the signals coming from all centre and surface coils were recorded during the same pulse. 

In order to measure the pulse of applied magnetic field, a coil with 100 turns of $150~\mu\mathrm{m}$ diameter wire was placed on the main rod at $1~\mathrm{cm}$ above the sample holder, where the magnetic influence of the sample can be neglected. A temperature sensor (Cernox probe) and a Hall probe (Arepoc LHP-MU) were also placed against the sample surface (as in Figure~\ref{Sample}). These three additional signals were acquired simultaneaously with the microcoil signals at $100~\mathrm{kHz}$.


The microcoils and the reference coil were calibrated by applying a magnetic pulse with the sample in its normal state ($T=100~\mathrm{K}$). The coil pick-up signals were then acquired simultaneously with the Hall probe signal during the application of the pulse; the Hall probe was used as a reference for the calibration.

Four successive magnetic pulses were applied to the sample. Two of them were applied at a temperature of $100~\mathrm{K}$ and served for the calibration of the centre and surface coils, respectively. The two other pulses were applied at $77~\mathrm{K}$ for the acquisition of the pick-up voltage of the $16$ centre and surface coils.

\section{Results and discussion}
\label{s:results}

We now turn to the description of the experimental results. We first discuss the time evolution of the magnetic flux density in the median plane of four given holes. On that basis, we compare the characteristics of the flux front propagation in the median plane and on the surface of the sample. We finally compare the trapped magnetic flux density in the median plane with that on the surface of the sample.

\subsection{Time dependence of the magnetic flux density in the hole}
\label{s:Time}

\begin{figure}[t]
\center
\includegraphics[width=10cm]{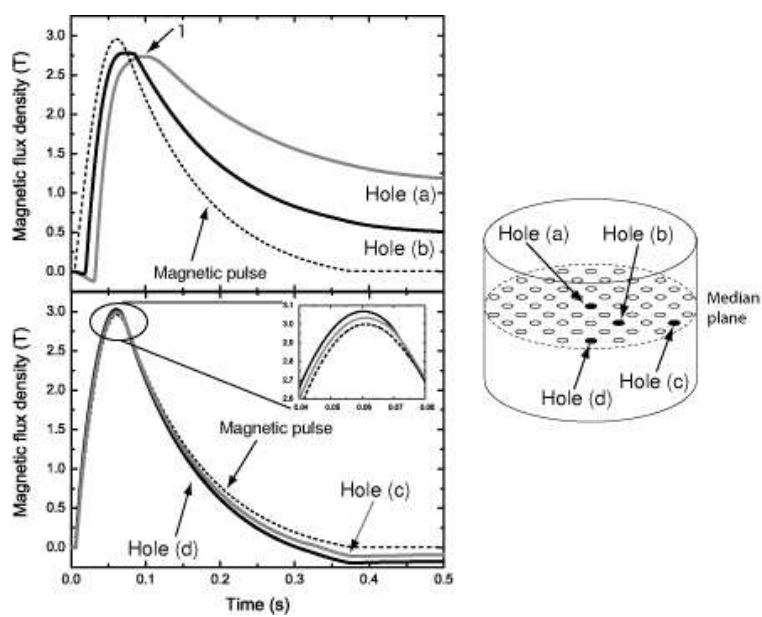}\caption{Time evolution of the magnetic flux density in the median plane of holes (a), (b), (c), and (d). The applied magnetic flux density is shown with a dashed line. }\label{FigureTimeEvol}
\end{figure}

We first consider the time evolution of the magnetic flux density in the median plane of the sample, {\it i.e.} for microcoils that are placed at an equal distance from the top and bottom surfaces. Four representative holes are investigated (cf. Figure~\ref{FigureTimeEvol}). The magnetic flux density in a given hole is calculated by integrating the pick-up voltage induced across the microcoils (which is proportional to $dB/dt$). In practice,  the pick-up voltage is biased by an unavoidable parasitic DC offset arising from the output of the instrumentation amplifier. The offset can be considered as remaining constant during the pulse ($<500~\mathrm{ms}$). To estimate this offset, the acquisition starts $5~\mathrm{ms}$ before the pulse is applied and the voltages recorded prior to the beginning of the pulse are averaged. The result is then subtracted from the raw data. 

Figure~\ref{FigureTimeEvol} shows the time evolution of the magnetic flux density, $B(t)$, in the median plane of the four representative holes (black and grey lines), together with the applied magnetic pulse (dashed line). Hole (a) is located at the centre of the sample, hole (b) is located at a half radius from the centre, and holes (c) and (d) are the closest to the sample outer edge as sketched on the right panel of Figure~\ref{FigureTimeEvol}. 

The time-dependence of the magnetic flux density in the holes has an overall pulse shape similar to that of the applied field, but however with a few different features. In hole (a), which joins the centres of the top and bottom faces, the pulse of flux density is preceded by a time interval during which $B(t)$ decreases slightly and is negative, {\it i.e.} the magnetic field in hole (a) is oriented in the opposite direction to the applied field. Thus, the $B(t)$ pulse is shifted in time with respect to the applied field and the magnetic flux density reaches a maximum after the applied field has already started to decay (see arrow 1 in Figure~\ref{FigureTimeEvol}). Moreover, when the applied field vanishes,  a positive trapped magnetic flux density remains in the hole.

In hole (b), during a short time-interval preceding the pulse, the magnetic flux density is also negative and thus oriented in the opposite direction to the applied field.  After a time delay that is shorter than for hole (a), the magnetic flux density in hole (b) increases with the applied field and reaches a maximum value when the applied field is maximum. Afterwards, it decreases with time and --- similarly to the situation in hole (a) -- a positive trapped magnetic flux density is measured in hole (b) at the end of the pulse.

In holes (c) and (d), located close to the sample edge, the behaviours of the magnetic flux density during the pulse are similar to each other. As soon as the applied field starts to increase, the magnetic flux density increases and reaches a maximum  that exceeds the applied field (see inset of Figure~\ref{FigureTimeEvol}). This behaviour, hereafter called {\it   flux concentration}, was also observed in AC regime where it was carefully checked not to  be due to a calibration artefact~\cite{Fluxinside}. The magnetic flux density in holes (c) and (d) remains larger than the applied flux density until the applied pulse starts decreasing. Then, it becomes smaller than the applied flux density. In particular, when the applied pulse vanishes, a negative trapped magnetic flux density is measured.

As often, simple approaches such as the Bean model allows one to   interpret the main features of the magnetization signal. Here however, the behavior predicted by the Bean model can be observed   in the time signal for hole (b) only, expect for the beginning of the   pulse.  The Bean model predicts a zero magnetic field in the hole until it is reached by the flux front. It then increases and remains smaller than the applied field. When the applied field reaches its maximum, the magnetic flux density in the hole saturates at a constant value because of the pinning of the vortices. At long times, it finally decreases down to a finite positive value that corresponds to the trapped field in the hole. The Bean model also predicts that both the saturation and the steady-state values depend on the critical current density and on the distance between the hole and the outer lateral surface.

Data clearly show deviations from the Bean picture. The Bean model fails reproducing, (i), the negative magnetic flux density at the beginning of the pulse in holes (a) and (b), (ii), the time delay between the maximum of the applied field and that of the field in hole (a), and (iii), the flux concentration and the negative trapped flux density in holes (c) and (d).

We suggest that these discrepancies may be attributed to demagnetizing effects appearing in short superconducting samples.
\begin{itemize}

\item[(i)] As the magnetic flux starts penetrating the sample from the edges, the magnetic field is found to be negative in holes (a) and (b). The sample being only partly penetrated, supercurrents flow in its outer region so as to screen its bulk from the applied  magnetic flux. The net negative induction which is observed in data suggests the occurrence of an {\it overscreening} mechanism, where the induced field exceeds the applied one. Such mechanism results from the demagnetizing effects that are associated with the geometry of the supercurrent trajectories. Overscreening has already been observed in flat rings, near their inner edge~\cite{Brandt2}, or in superconducting tubes, close to the inner wall~\cite{Sam}. In the present case, the overscreening mechanism is likely to be strengthened by the rather complex trajectories of the supercurrents, which must flow around the holes. 

\item[(ii)] Right after the applied field has reached its   maximum amplitude and starts decreasing, the magnetic field in hole   (a) keeps increasing. At that instant, the supercurrents   flowing in the outer part of the sample must reverse their direction   with respect to that of the supercurrents flowing in the remaining   part of the sample.  The resulting demagnetizing field through hole   (a) is now oriented in the same direction as the applied field and   thus reinforces the flux density in that hole. As a result, the occurrence of the maximum in the magnetic flux threading hole (a) is delayed.

\item[(iii)] During the increasing part of the applied pulse, a flux   concentration is observed in holes (c) and (d). As mentioned above,   the demagnetizing field produced in the central hole during that   time is oriented in the direction opposite to the applied field. The   returning lines of the demagnetizing field, however, are oriented   in the same direction as the applied field. It is therefore likely   that the periphery of the sample is crossed by the returning lines of   that demagnetizing field; in particular, holes (c) and (d) intercept   such lines. Since these lines remain after the applied   field has vanished, the remnant magnetic flux density in the   holes located near the lateral surface of the sample is negative, as   indeed observed in Figure~\ref{FigureTimeEvol}.

\end{itemize}
\begin{figure}[t]
\center
\includegraphics[width=8cm]{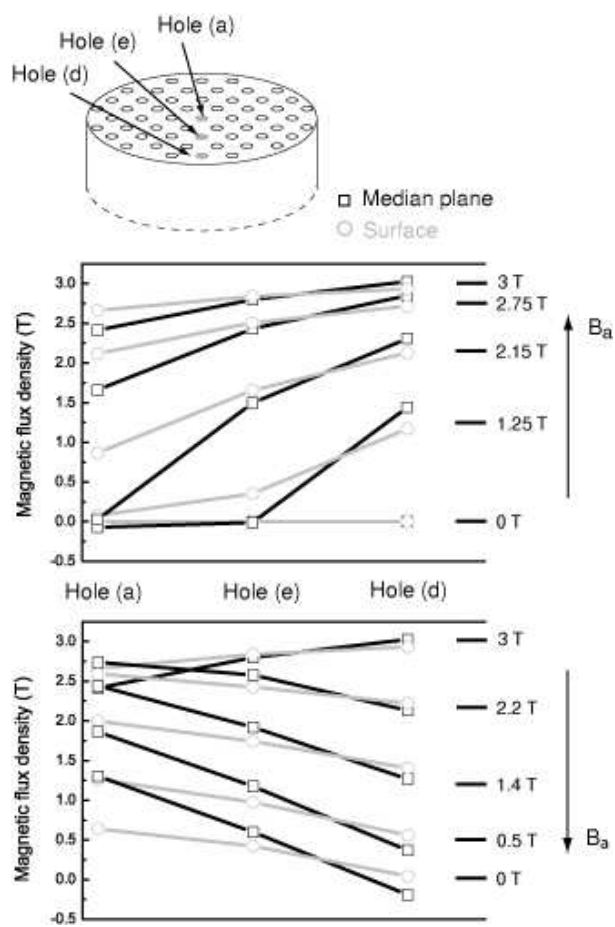}\caption{Magnetic flux   density in holes (a), (e), and (d), whose positions are represented   on the top panel. The flux density is shown in the median plane and   on the sample surface, for both the increasing part (middle panel)   and the decreasing part of the pulse (bottom   panel).}\label{FigureLinescan}
\end{figure}

In order to study whether the demagnetizing field  influences the magnetic flux density similarly on the surface and in the bulk, let us consider three successive holes along a given radial line. We reproduce in Figure~\ref{FigureLinescan} the magnetic flux density in the median plane and on the surface of these holes, namely (a), (e), and (d), at several consecutive times during the pulse. During the increasing part of the pulse (Figure~\ref{FigureLinescan} - middle panel), the magnetic flux density is larger at the sample periphery than near the sample centre. The spatial decay of the magnetic flux density along the radial line is stronger in the median plane than at the sample surface, due to demagnetizing effects. For hole (d), which is the closest to the sample edge, we observe a flux concentration in the median plane. This behaviour is not observed near the surface. Similarly, for hole (a), the magnetic flux density is found to be negative at the beginning of the pulse in the median plane, whereas it is always positive near the surface. 

As the pulse starts decreasing (Figure~\ref{FigureLinescan} - bottom panel), the magnetic flux density in the median plane of the central hole (a) first increases before decreasing, whereas the flux density on the surface is found to decrease monotonically. Once the applied field vanishes, the trapped flux density in the centre of the sample is always larger in the median plane than near the surface. Moreover, the spatial variation of the magnetic flux density along the radial line is also smaller near the surface than in the median plane, because of the demagnetizing fields. In particular, the magnetic flux density is found negative in the median plane of the peripheral hole, whereas it remains positive at the sample surface. 

It is therefore observed that the effects attributed to the demagnetizing field are different on the surface than in the median plane. Near the surface, the demagnetizing field cause the spatial variations of $B_z$ along the radial line to differ from that in the median plane. In the median plane of the sample, it is responsible for the effects (i), (ii), and (iii) which were described above. These effects are not observed on the surface.

\subsection{Characterisation of the propagation of the flux front in the sample}
\label{s:Fluxpropagation}

On the basis of the time evolution of the magnetic flux density in each of the 16 holes,  we now characterize the propagation of the flux front during the pulse in both the median plane and on the surface. The flux front is defined as the boundary between the regions that are already penetrated by the magnetic flux and those that are not. In practice, the characterization of that boundary requires defining a threshold for the magnetic flux density above which a given location can be considered as being penetrated. In the following, we consider a given hole as penetrated if 
\begin{equation}
B_{in}\geq\frac{B_a}{10},
\end{equation}
 where $B_{in}$ stands for the magnetic flux density in the hole and $B_a$ is the applied magnetic flux density.  Accordingly, we define for each hole a penetration time, $t_{pen}$, as the time $t=t_{pen}$ at which $B_{in}(t_{pen})=B_a(t_{pen})/10$.

\begin{figure}[t]
\center
\includegraphics[width=10cm]{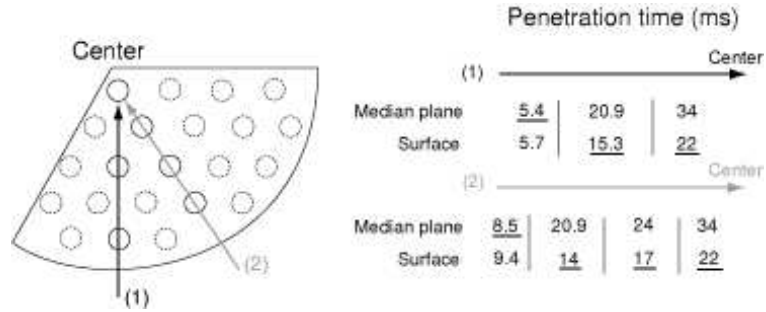}\caption{Penetration times of the flux front in the median plane and on the surface of 6 holes belonging to one of the radial lines (1) and (2). For each hole, the shortest penetration time (median plane or surface) is underlined.}\label{FigurePenetrationTime}
\end{figure}

The penetration times of holes belonging to two different sample radial lines is shown in Figure~\ref{FigurePenetrationTime}. The penetration time of a given hole is reported in the median plane and on the surface; the shortest penetration time is underlined. Along the radial line (1), the flux front first penetrates the peripheral hole in its median plane ($t_{pen}=5.4~\mathrm{ms}$) before reaching its surface ($t_{pen}=5.7~\mathrm{ms}$). In the other holes along that line, the situation is reversed: the flux penetration first occurs on the surface. Along the radial line (2), we observe a similar behaviour. The peripheral hole is penetrated in its median plane first whereas we observe the opposite situation in the other holes. 

Two observations may be drawn from the results presented in Figure~\ref{FigurePenetrationTime}:
\begin{itemize}
\item[(1)] close to the sample edge, the flux front moves faster in the median plane than on the surface,
\item[(2)] as the flux front moves further toward the sample centre, it proceeds at a higher speed on the surface than in the bulk. 
\end{itemize}

\begin{figure}[t]
\center
\includegraphics[width=10cm]{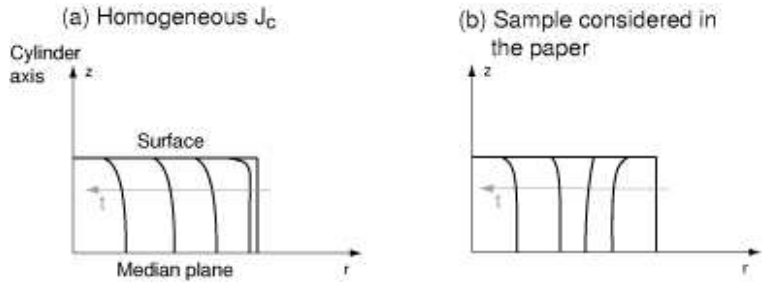}\caption{Sketch of the propagation of the flux front in the $(r,z)$ plane of a cylindrical sample, at different times during the ascending part of the pulse. Two situations are depicted: (a) a sample having a uniform critical current density, and (b), the sample considered in the present work. }\label{FigureFrontPropagation}
\end{figure}

Figure~\ref{FigureFrontPropagation}-(a) shows a sketch of the flux front at different times in the $(r,z)$ plane, for two situations: (a) - left panel - a sample with a uniform critical current density flowing along the azimuthal direction, and (b) - right panel - the sample considered in this paper, on the basis of the above results. The qualitative description of the flux front propagation in panel (a) is based on the results of Brandt in Ref.~\cite{Brandt} for a cylindrical sample with the same aspect ratio as that of sample (b). In sample (a), the flux front systematically moves faster on the surface than in the median plane. By contrast, in the sample considered in the paper, the flux front proceeds at a faster pace in the median plane for the largest radial positions, whereas it is faster on the surface for the lowest radial positions.

The properties of the flux front propagation are closely related to the detailed distribution of the critical current density in the sample. A large critical current density --- a strong pinning --- hampers the penetration of the flux. Thus, Figure~\ref{FigureFrontPropagation} suggests that the critical current density is not uniform in the sample. In particular, it has a lower value in the median plane, close to the sample edge where the flux front moves faster than on the surface. Such non-uniformities in the critical current density may arise either (i) from intrinsic superconducting properties, $J_c(r,z)$~\cite{Dewhurst, Lo}, or (ii), from a non-uniform heat production in the sample and a non-uniform surface heat exchange coefficient, $J_c(T)$~\cite{Fujishiro}. 

\subsection{Trapped magnetic flux in the volume of the sample}
\label{s:Trapped}

We now turn to the analysis of the spatial distribution of the trapped field in the median plane and on the surface of the sample. 

\begin{figure}[t]
\center
\includegraphics[width=10cm]{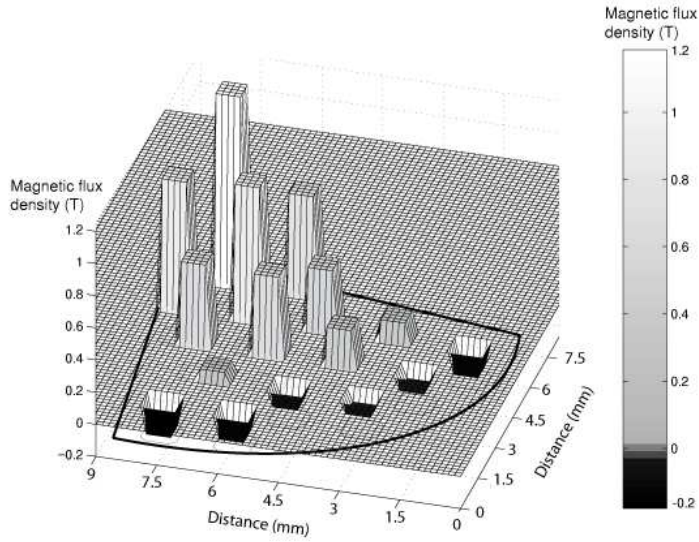}\caption{Distribution of the trapped magnetic flux density in the median plane of the holes shortly after the end of the pulse, at $t=0.5~\mathrm{s}$.  }\label{FigureTrapped}
\end{figure}

Figure~\ref{FigureTrapped} shows the magnetic flux density in the median plane of the 16 holes considered in this paper, measured shortly after the end of the pulse, $t=500~\mathrm{ms}$. The holes are represented by a square tower whose height is proportional to the trapped flux density. The trapped flux density is found to be maximum at the centre of the sample, $B^{\mathrm{median~plane}}_{\mathrm{trapped}}=1.2~\mathrm{T}$, and is smaller in holes that are located further from the centre. It is also observed that the peripheral holes have a negative trapped flux density in their median plane. 

The maximum trapped flux density measured in the median plane of the sample allows us to estimate an average critical current density, $\tilde{J_c}$, flowing in the sample. If we consider as a first approximation that the holes are negligible obstacles for the current flow, the trapped magnetic flux density at the centre of the median plane of a cylinder with a radius $R$ and a height $L$ is given by~\cite{Chen},
\begin{equation}\label{avJc1}
B_{\mathrm{trapped,max}}=\mu_0\tilde{J_c}\frac{L}{2}\,\ln\frac{R+\sqrt{R^2+L^2/4}}{L/2}. \end{equation}
From the field density values of Figure~\ref{FigureTrapped}, we thus find
\begin{equation}\label{avJc1-1}
\tilde{J_c}\sim1.5~10^8~\mathrm{A/m^2}
\end{equation}
as a first approximation.

However, the neglect of the holes may lead to non-negligible errors, as shown in previous works~\cite{InfluenceHoles, FEM}.  Following   the same analysis as in~\cite{InfluenceHoles, FEM}, we find that the   maximum trapped flux density in the drilled sample is expected to be   about $40~\%$ smaller than that in a sample without holes, assuming   the same uniform critical current density. Taking this corrective factor into account yields
\begin{equation}\label{Jccenter}
\tilde{J_c}\sim2.5~10^8~\mathrm{A/m^2}
\label{Jc-second}
\end{equation}
as a better estimate.

The trapped magnetic flux density measured on the surface of the hole differs from that measured in the median plane. At the central hole, we measured $B^{\mathrm{surface}}_{\mathrm{trapped}}=545~\mathrm{mT}$ (data shown in Figure~\ref{FigureLinescan}). This value is smaller than that in the median plane by a factor $B_{\mathrm{trapped}}^{\mathrm{median~plane}}/B_{\mathrm{trapped}}^{\mathrm{surface}} \approx 1.18/0.545 \approx 2.17$.

This value can be compared to elementary models. An analytical   expression of the magnetic flux density along the central axis of   the sample has been calculated in the case of a cylindrical sample without holes, see Ref.~\cite{Chen}.  Given a radius, $R$, and a height, $L$, and assuming a uniform critical current density, $J_c$, the magnetic flux density along the axis of the cylinder is given as
\begin{eqnarray}
\label{Bz}
B_z(z) & = &\frac{1}{2}\mu_0J_c\left\{\left(z+\frac{L}{2}\right)\ln\left(\frac{R+\sqrt{R^2+(z+L/2)^2}}{|z+L/2|}\right)\right.\nonumber\\
& & \left.-\left(z-\frac{L}{2}\right)\ln\left(\frac{R+\sqrt{R^2+(z-L/2)^2}}{|z-L/2|}\right)\right\},\label{analytic}
\end{eqnarray}
where $z$ denotes the elevation from the center. According to this relation, for a plain sample (i.e. containing no holes) with the dimensions of the sample described in Section~\ref{ss:sample} ($L=10~\mathrm{mm}$ and $R=8~\mathrm{mm}$), the ratio between the maximum trapped flux in the median plane and on the surface is equal to
\begin{equation}
\lim_{z\rightarrow L/2} \left.\frac{B_z(0)}{B_z(L/2)}\right|_{L/R=1.25}=1.7~.\label{ratio}
\end{equation}

\begin{figure}[t]
\center
\includegraphics[width=8cm]{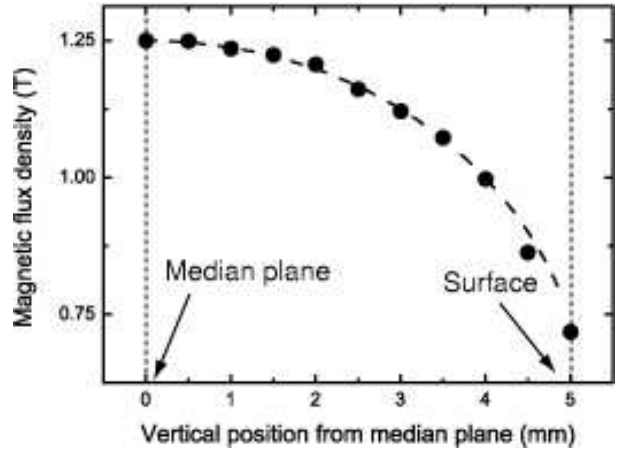}\caption{3D finite-element simulation of the vertical component of the magnetic flux density as a function of the vertical coordinate, $B_z(z)$, in the central hole of the drilled sample considered in the paper (filled circles). The analytical expression of $B_z(z)$ at the centre of a sample without holes is shown with a dashed line.}\label{FigureBz}
\end{figure}

Since the holes modify the current stream lines, there is {\it a   priori} no reason that equation~(\ref{analytic}) remains valid for a drilled sample. To better understand the influence of the holes on the magnetic induction along the axis of the sample, we performed a 3D numerical simulation with the finite-element method described in~\cite{FEM} and calculated the trapped flux density for the drilled sample considered in this study. Accordingly to our previous   estimate for the critical density, equation~(\ref{Jc-second}), we   assume a uniform and field-independent critical current density   $J_c=2.5~10^{8}~\mathrm{A/m}^2$. The resulting magnetic flux   density is shown as a function of the vertical position from   the median plane in Figure~\ref{FigureBz}-(b), for the central hole   of the sample. The analytical expression for $B_z(z)$,   equation~(\ref{analytic}), is shown with a dashed line. It can be   seen that the dependence of $B_z(z)$ of the drilled sample is very   close to that of the plain sample. In particular, the ratio between   the maximum trapped flux in the median plane and on the surface is   found to be equal to $\approx 1.74$ in both cases.

This simulated ratio, obtained for a sample with a uniform critical current density, is different from the ratio measured in our drilled sample (for which we found $2.17$). The difference may result from two causes. First, the position of the microcoil inside the hole is not perfectly known and the data is obtained from an average over the microcoil volume. As a small variation in $z$ close to the surface of the sample leads to a large change in $B(z)$ (as observed in Figure~\ref{FigureBz}), the measured ratio can easily differ from the simulated one. Second, the sample may exhibit a non-uniform distribution of its critical current density, such effect was not taken into account in the numerical model.

\section{Conclusions}
\label{s:Conclusion}

We use the holes in a drilled YBaCuO cylinder to probe the local magnetic flux density in the median plane and on the surface of the sample, during a pulsed-field magnetization. As the magnetic pulse penetrates the sample, the induced emfs are simultaneously measured across a series of microcoils that are placed in the median plane or close to the surface of the sample. This technique enables us, for the first time, to characterize of the flux propagation in the bulk of the sample, instead of a few millimeters above its surface as is the case with conventional field mapping techniques. 

We analyse the time evolution of the magnetic flux density in the holes (median plane and surface) and show that some holes may serve as a returning path for the demagnetizing field, yielding a different behaviour from what is predicted by simple approaches such as the Bean model. In particular, during the ascending part of the pulse, we observe a flux concentration inside the holes that are close to the sample edge and a negative magnetic flux density in the holes near the centre of the sample. When the pulse is over, the trapped magnetic flux density is negative in the holes that are close to the sample edge whereas it is positive at the centre of the sample. These demagnetizing effects, associated with the returning lines, are found to be stronger in the median plane than on the surface of the sample.

In order to further characterize the flux propagation in the sample, we define in each hole the required time for the external flux to reach this hole. We find that holes located at the sample edge are penetrated in their median plane sooner than at their surface, while the opposite is found in the other holes. These observations may arise from either non-uniform superconducting properties in the sample or from particular thermal boundary conditions at the surface of the sample.

For the first time, we show experimentally that the trapped flux density is larger in the median plane than on the surface. This finding is confirmed by modelling results. Moreover, the measured trapped magnetic flux density in the centre of the median plane of the sample allows us to directly estimate the critical current density in the volume of the sample. Conventional methods, on the other hand, suffers from uncertainties in the distance between the magnetic probe and the surface of the sample. With the help of numerical modelling, we compare the magnetic flux density in the median plane and on the surface of the sample. In particular, we show that the measured ratio between the maximum trapped flux in the median plane and that on the surface slightly differs from the ratio predicted by numerical results.

\section{Acknowledgments}

This work was funded in part by the contract FP6 {\it Structuring the European Research Area, Research Infrastructure Action} (EuroMagNet, contract RII3-CT2004-506239). We also thank the \textit{Fonds de la Recherche Scientifique (FRS-FNRS)} of Belgium and the University of Li\`ege for cryofluid and equipment grants.

\end{document}